%% file: main.tex
\documentclass[submission]{eptcs}
\providecommand{\event}{GaM 2016} 
\usepackage{breakurl}

\usepackage[pdftex]{graphicx}
\usepackage{paralist}
\usepackage{amsmath}
\usepackage{wrapfig}
\usepackage{amsthm}
\usepackage{subcaption}
\usepackage{array}
\usepackage{tikz}
\usetikzlibrary{calc}

\usepackage[]{hyperref}
\hypersetup{
  colorlinks=true, linkcolor=black, citecolor=black, urlcolor=black,
}

\theoremstyle{definition}

\newcommand{\dno}{\emph{deleteNextObject()}}

\title{Towards a Step Semantics \\for Story-Driven Modelling}
\author{
G{\'e}za Kulcs{\'a}r 
\institute{ 
Technische Universit\"at Darmstadt\\
\email{geza.kulcsar@es.tu-darmstadt.de} 
}
\and
Anthony Anjorin
\institute{
Chalmers~$|$~University of Gothenburg\\
\email{anjorin@chalmers.se}
}
}
\def\titlerunning{Towards a Step Semantics for Story-Driven Modelling}
\def\authorrunning{G{\'e}za Kulcs{\'a}r \& Anthony Anjorin}

\begin{document}
\maketitle 

\begin{abstract}
Graph Transformation (GraTra) provides a formal, declarative means of specifying model transformation.
In practice, GraTra rule applications are often \emph{programmed} via an additional language with which the order of rule applications can be suitably controlled.

Story-Driven Modelling (SDM) is a dialect of programmed GraTra, originally developed as part of the Fujaba CASE tool suite.
Using an intuitive, UML-inspired visual syntax, SDM provides usual imperative control flow constructs such as \emph{sequences}, \emph{conditionals} and \emph{loops} that are fairly simple, but whose interaction with individual GraTra rules is nonetheless non-trivial.
In this paper, we present the first results of our ongoing work towards providing a formal \emph{step} semantics for SDM, which focuses on the \emph{execution} of an SDM specification.
\end{abstract}

\input{introduction}
\input{relwork}
\input{background}
\input{semantics}
\input{conclusion}

\bibliographystyle{eptcs}
\bibliography{pubs,geza}

\end{document}

%% file: introduction.tex
\section{Introduction and Motivation}
\label{sec:intro}

Graph Transformation (GraTra) provides a formal, declarative means of specifying how graph-like structures can be manipulated and changed.
This is useful in numerous application domains including the specification of model transformations, a central task in Model-Driven Engineering~(MDE)~\cite{MDE}.
Although GraTra rules can simply be applied as long and as often as they are applicable, in practice more control over rule application is often required for complex transformations.   
This can be achieved in many ways including grouping and ordering rules in layers~\cite{DBLP:conf/rta/LoweB93}, or by providing a dedicated additional language, with which GraTra rules can be suitably ``programmed"~\cite{Henshin2010,DBLP:journals/sttt/GhamarianMRZZ12,Schurr1999,Zundorf2002}.

Story-Driven Modelling (SDM)~\cite{Zundorf2002} is a dialect of programmed GraTra introduced by Z\"undorf and released as part of the Fujaba CASE tool suite.
Originally inspired by PROGRES~\cite{Schurr1999}, SDM combines object-oriented concepts and GraTra rules in a single, formal and integrated language, with an intuitive, UML-inspired visual concrete syntax.
To provide control over rule application, SDM introduces usual imperative control flow constructs including sequences and conditionals.
These constructs are relatively simple, but their interaction with the embedded, normal GraTra rules is nonetheless non-trivial and involves a series of careful design decisions.

The current formalisation of SDM semantics by Z\"undorf~\cite{Zundorf2002} is based on pairs of graphs representing input-output pairs of GraTra rule applications, first defined for single GraTra rules and then extended to combinations of rule applications, programmed by an iterative control flow.  
This denotational style of formalisation is useful to determine the correctness of results of rule applications. 
Nevertheless, although Z\"undorf mentions typical optimization techniques for SDMs such as binding variable names to input model elements to reduce the search space of subsequent GraTra rule applications, the exact semantics regarding the combination of conditionals and such bindings remains largely unspecified.
Such a ``high-level'' semantics can be advantageous from a specification point of view, but for GraTra researchers and tool developers aiming to provide, e.g., static analyses for SDM specifications, a more detailed \emph{execution-based} semantics is also required as the set of output models is not available for every possible input.

In this paper, therefore, we propose to complement \cite{Zundorf2002} with an operational semantics for SDM that focuses on the control flow and execution of an SDM specification. 
To this end, we define a \emph{step semantics} whose step concept relies on GraTra rule applications, ensuring that the resulting formalisation is particularly comprehensible for our target audience: researchers and tool developers in the GraTra community. 
We demonstrate this on a fundamental set of SDM constructs, which we shall refer to as \emph{basic SDM}, consisting of single rule applications, sequences, conditionals and head-controlled loops. 
We define the set of syntactically valid combinations of these constructs by means of a graph grammar that generates exactly the valid control flows of basic SDM. 
Based on this, we provide a further set of GraTra rules whose application yields a step concept for our semantics.

%% file: relwork.tex
\section{Related Work}
\label{sec:relwork}

The initial, and to the best of our knowledge, only formal SDM specification is provided by Z\"undorf in \cite{Zundorf2002} and is based on the semantics for PROGRES~\cite{Schurr1999}.
Technical reports on SDM such as \cite{vDHP12b1} go a long way towards clarifying SDM syntax and semantics and are probably more readable and accessible to end users than any formal specification.
For researchers and tool developers, however, whose aim is to provide new tool support or extend the language, such semi-formal documentation is not precise enough, especially concerning control flow constructs and their interaction with embedded GraTra rules. 

\medskip
\noindent\textbf{Formalisation techniques.}
Dynamic Meta-Modelling (DMM) is introduced by \cite{DBLP:conf/uml/EngelsHHS00} as a visual approach to specify the dynamic behaviour of (visual) languages.
It is closely related to and can be viewed as a generalisation of Graphical Operational Semantics (GOS)~\cite{DBLP:conf/icalp/CorradiniHM00}.
Both approaches are inspired by Structural Operational Semantics (SOS)~\cite{DBLP:journals/jlp/Plotkin04a} and they employ GraTra rules to induce a transition system representing the specified semantics.
These techniques have been applied in \cite{DBLP:conf/fase/HeckelS01,DBLP:journals/entcs/HeckelZ01} to formalise UML collaboration diagrams and the pattern language used in Fujaba.
In this paper, we focus more on the control flow constructs of SDM and less on the embedded patterns, which we restrict to normal GraTra rules without integrating advanced object-oriented constructs.

SDM specifications show some resemblance to UML activity diagrams, for which an executable step semantics has been proposed in~\cite{Knieke04}, inspired by a similar approach to a statechart semantics~\cite{statechart}. 
The DMM approach has been applied to the semantics of activity diagrams in~\cite{StorrleH05}, where the semantics of Petri-nets is taken as an initial point for comparison. 
The authors conclude that despite superficial similarities, the intention of higher-level UML constructs notably diverges from a Petri-net semantics. 
Although these approaches are similar to our token and step concepts, nondeterministic pattern matching, failure as a branching condition, as well as variable binding are not considered directly in the case of activity diagrams (as Petri-nets are also missing analogous concepts). 
Moreover, the formalisms of~\cite{statechart,Knieke04} do not involve GraTra concepts (even if an encoding into GraTra would be possible). 
An analysis of activity diagrams based on GraTra has been proposed in~\cite{adgratra2} involving an object flow concept and a rule-based semantics. 
Nevertheless, in contrast to our approach concentrating on execution, the proposed semantics focuses on the resulting graphs and the properties of the involved GraTra rules.

As demonstrated by \cite{Zundorf2002}, it is possible to provide a denotational semantics for SDM.
This can be achieved by mapping an SDM specification to the set of all pairs of possible input and output graphs.
While this can be viewed as an elegant and compact formalisation, we believe that a complementary operational approach that uses GraTra along the lines of \cite{DBLP:conf/icalp/CorradiniHM00,DBLP:conf/uml/EngelsHHS00} is  crucial for further research and development on SDM.
Similar arguments apply to other approaches including, for example, abstract state machines (cf. \cite{DBLP:journals/apal/Borger05} for a comparative study). 

\medskip
\noindent\textbf{Alternatives to SDM.}
Existing and established GraTra tools take different approaches to enabling programmed GraTra.
PROGRES~\cite{Schurr1999} and Viatra~\cite{DBLP:conf/sac/BaloghV06}, similarly to SDM, provide a dedicated control flow language, rich and complex enough to completely replace any host language.
Approaches such as EMF-IncQuery~\cite{UjhelyiBHHIRSV15} and Groove~\cite{DBLP:journals/sttt/GhamarianMRZZ12} have chosen to concentrate fully on providing a rich pattern language (in the case of EMF-IncQuery even without side effects) for GraTra.
These approaches rely on a host language (Java, Prolog, Xtend) for cases where users wish to additionally control GraTra rules.
Approaches such as Henshin~\cite{Henshin2010} are in-between, focussing primarily on patterns but still providing a relatively simple and high-level language such as \emph{transformation units}~\cite{DBLP:conf/birthday/KreowskiKR08} to program individual GraTra rules.

%% file: background.tex
\section{Preliminaries}
\label{sec:background}
\label{sec:gratrabasics}
\label{sec:sdmbasics}

In this section, we recapitulate the basics of GraTra based on~\cite{Fundamentals,Handbook}, and briefly recall the motivation and background of SDM, introducing standard terminology required for the paper with a simple example.

\medskip
\noindent\textbf{Graph transformation.} 
Graphs are ubiquitously used for capturing different structures in various domains as they are mathematically tractable but still comprehensible and can be easily visualised. 
In this paper, a graph is defined as a tuple $G = (V_G,E_G,src_{G}:E_G \rightarrow V_G,trg_{G}:E_G \rightarrow V_G)$, $V_G$ being the set of \emph{nodes}, $E_G$ the set of \emph{edges}, $src_{G}$ and $trg_{G}$ the source and target functions, assigning edges to their source and target nodes, respectively.
Given graphs $G$ and $H$, a \emph{graph morphism} $m: G \rightarrow H$ is a pair of functions $m_V: V_G \rightarrow V_H$ and $m_E: E_G \rightarrow E_H$ preserving connectivity, i.e., $m_V(src_{G}(e_G)) = src_{H}(m_E(e_G))$ for each $e_G \in E_G$ and analogously for target functions. 
A morphism is \emph{injective} if both functions are injective.

\emph{Type graphs} are often used to enhance graphs with additional structure by assigning types to nodes and edges. A type graph is a distinguished graph $TG$. 
A \emph{typed graph (G, $t$) over TG} is a graph $G$ together with a graph morphism $t: G \rightarrow TG$. 
A \emph{typed graph morphism} is a morphism $m: (G,t) \rightarrow (H,t')$ preserving typing, i.e., $t_V'(m_V(v_G)) = t_V(v_G)$ and $t_E'(m_E(e_G)) = t_E(e_G)$ for each $v_G \in V_G$ and $e_G \in E_G$.

In this paper, we use the Single Pushout (SPO) approach to GraTra, which is formalised in a categorical framework (cf.~\cite{Handbook} for formal details). 
The SPO approach to GraTra utilizes the notion of a \emph{partial graph morphism}. 
A partial graph morphism $p: G \rightharpoonup H$ is a graph morphism $dom(p) \rightarrow H$, where $dom(p)$ is a sub-graph of $G$. 
The definition can be extended to the notion of \emph{partial typed graph morphism} analogously to typed graph morphisms.

\begin{wrapfigure}[5]{hR}{0.12\textwidth}
	\vspace{-0.3cm}
	\input{po.tikz}
\end{wrapfigure}
A \emph{graph transformation rule} (or simply \emph{rule}) is a partial graph morphism $r: L \rightharpoonup R$ where the graphs $L$ and $R$ are also called the \emph{left-hand side} and the \emph{right-hand side} of the rule $r$, respectively. A \emph{match} of this rule in a graph $G$ is a total injective morphism $m: L \rightarrow G$.  The set of valid matches can be further constrained by \emph{negative application conditions} (NAC). 
A NAC is a morphism $n: N \rightarrow L$ which forbids a match $m$ if there is an image of $N$ in $G$ via $m$. 
The \emph{application of a rule} $r$ at match $m$ in graph $G$, denoted as $G \overset{r,m}{\Longrightarrow} H$ (or simply $G \overset{r}{\Longrightarrow} H$ if the match is not relevant), transforms $G$ to the graph $H$ as seen in the diagram to the right, also called a \emph{pushout} (PO) diagram. 
Intuitively, the pushout diagram defines rule application as follows: the images $m(L\setminus dom(r))$ of elements not in the domain of $r$ are deleted from $G$ and the elements $m'(R \setminus r(L))$ not in the image of $L$ in $R$ are added to $G$, yielding the graph $H$. Again, the generalisation to the typed case is straightforward. Practically, the side effect of using SPO rules (compared to the Double Pushout approach) is that if nodes are deleted by a rule application, their incident edges are also deleted, even if they are not explicitly in the deletion sub-graph derived from $r$.
A graph grammar $GG = (G_S,\mathsf{R})$ consists of a \emph{start graph} $G_S$ and a set of graph transformation rules $\mathsf{R}$. 
The \emph{language generated by GG} contains those and only those graphs which can be reached via a rule application or a chain of rule applications from the start graph, i.e., is a set of graphs $L_{GG} = \{H ~|~ \exists G_S \overset{r_1}{\Longrightarrow} ... \overset{r_n}{\Longrightarrow} H\}$ where $r_1, ..., r_n \in \mathsf{R}$ and $n \geq 0$. 

\medskip
\noindent\textbf{Story-Driven Modelling.}
Story-Driven Modelling (SDM) has been originally proposed as a dialect of programmed GraTra, which enhances declarative GraTra rules with a control flow, describing the application order of the rules and including basic imperative control structures such as conditionals (if-then-else) and for-each loops. 

In this paper, an \emph{SDM specification}, referred to as a \emph{story diagram}, consists of a \emph{control flow graph} and a mapping of control flow nodes to \emph{story patterns}. 
A control flow graph is a graph consisting of control flow nodes and edges defining the imperative control structure. 
A story pattern is a typed GraTra rule as defined previously. 
We say that a story pattern is contained in a control flow node if the latter is mapped to the former.
In this case, the control flow node is referred to as a \emph{story node}.

SDM execution starts at a special control flow node, referred to as a \emph{start node} and continues along the edges of the control flow graph according to the semantic rules of SDM, which we shall define in this paper.
\emph{Story node execution} is formalized as an attempt to apply its contained story pattern (which is a GraTra rule) on the underlying input graph, which can either be successful (if the rule is applicable) or fail. 
SDM execution terminates if it arrives at a special control flow node referred to as a \emph{stop node}, or if the execution of some story pattern fails unexpectedly (what this  exactly means will be formalised by our semantic rules).
A story diagram is thus a program that takes as input a graph $G$ and outputs another graph $G'$, which can be derived from $G$ via a chain of GraTra rule applications $G \overset{r_1,m_1}{\Longrightarrow} G_1 \overset{r_2,m_2}{\Longrightarrow} G_2 \Longrightarrow ... \Longrightarrow G'$ that is compatible with the specified control flow graph of the story diagram.

\begin{figure}[!htbp]
	\includegraphics[width=\textwidth]{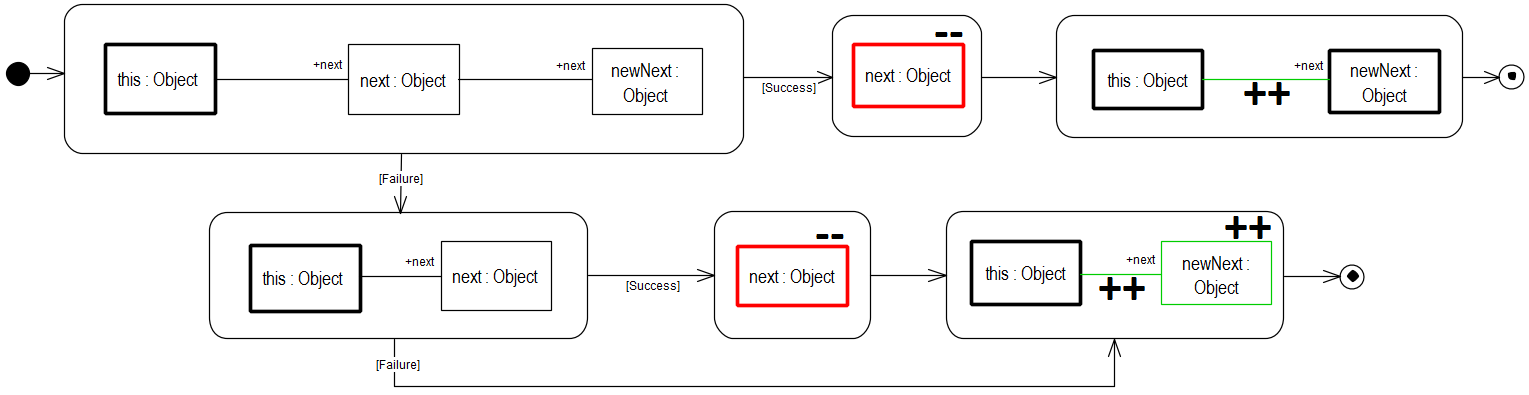}
	\caption{Example SDM method specification: \dno}
	\label{fig:example_dno3}
\end{figure}

\medskip
\noindent\textbf{Example.} An \emph{ordered list} consists of nodes (objects) linked to each other via single \emph{next} edges (links), where the next reference of the last object in the list is undefined (null).
Figure~\ref{fig:example_dno3} depicts the visual specification of a simple operation called \dno\ as a story diagram, which deletes a node from a list while retaining the linked structure of the list. It also guarantees that the current list object is not the last one in the list after execution.

A story pattern $r: L \rightharpoonup R$ is represented compactly by merging $L$ and $R$ as follows: black elements constitute $L \cap R$ and are retained, red elements with a ``-{}-" markup constitute $L \setminus R$ and are deleted, while green elements with a ``++" markup constitute $R \setminus L$ and are created by the rule.   
Finally, note that the first story pattern of \dno\ has a partial match already consisting of a node \emph{this} that is mapped to the object on which \dno\, is invoked, in the usual object-oriented fashion.  
Such nodes are referred to as \emph{bound} variables and are denoted with a bold border.

The story diagram consists of a \emph{start node} (filled circle, on the left), two \emph{stop nodes} (circle with a dot, on the right) and \emph{story nodes} in between. 
The execution of the story diagram starts at the start node, follows the first control flow edge (depicted as an arrow) leading to a story node which is a conditional. It has two outgoing arrows, one labelled with \emph{Success} and one with \emph{Failure}. 

If the story pattern in the conditional node is successfully matched, the control is passed over along the \emph{Success} arrow, otherwise along \emph{Failure}. 
Note that the two branches of this conditional never meet again and they end in different stop nodes. The \emph{Failure} branch (bottom row) starts with a conditional again (thus demonstrating a nested conditional), but is different from the first one as the branches join again in the story node just before the bottom stop node. 
If execution arrives at the second conditional, it means that the current object is followed by at most one object. In the second conditional, we check if at least one \emph{next} object exists. 
Note that it is allowed to have the variable \texttt{next} as unbound again as we know that the binding in the first conditional was unsuccessful. 
If we manage to match at least this smaller structure, then we follow the \emph{Success} arrow, deleting \texttt{next} and proceeding to the last story node. 
If \texttt{this} has no \emph{next} object at all, we skip to the last story node along \emph{Failure}. 
Independent of the executed branch, \texttt{this} now stands without a follower---which we create in the last control flow node.

\medskip
\noindent\textbf{Denotational SDM semantics.} In~\cite{Zundorf2002}, Z\"undorf defined an SDM semantics for the Fujaba tool suite, inspired by and partly based on the PROGRES approach~\cite{Schurr1999}. The basic element of this semantics is a single story node, which serves as a basis for the semantics of compound structures (sequences, conditionals, loops). In this denotational approach, the semantics of story node execution is given in terms of pattern matching in an input graph. Particularly, the semantics $Sem(n)$ of a story node $n$ containing GraTra rule $r$ is a set of graph pairs $Sem(n) = \{(G,H) ~|~ G \overset{r}{\Longrightarrow} H \}$. Thus, the semantics of a single step is based on GraTra rule applications which corresponds to our approach. Note that in the present paper, we do not explicitly specify the semantics of a single pattern matching step in terms of attaching input objects to the variables. In this respect, our technique is analogous to the one in~\cite{Zundorf2002} where it has been thoroughly specified. Through the denotational semantic style, SDM remains closely connected to GraTra theory, the resulting semantic domain is simple, and can be effectively used for testing an SDM method against, e.g., pairs of input-output models in a model transformation scenario.

In some cases, however, this approach might not capture the required level of detail. 
For instance, in applications of programmed GraTra in general and SDM in particular, the intended behaviour of an SDM method is mostly that in case a rule cannot be applied, the execution terminates immediately and optionally, the user gets informed about which pattern failed to match. The handling of termination requires an operational viewpoint and is not captured by the denotational approach. 

Particularly, as a consequence and crucial difference, the semantics according to \cite{Zundorf2002} of story node $n$ with rule $r$ and input graph $G$ is $Sem(n) = \{(G,G)\}$ if $r$ is not applicable to $G$.
In a sequential composition of story nodes starting with story node $n$, therefore, this means that the input graph remains unchanged after $n$, but leaves open the possibility of executing further story nodes in the sequence after $n$.

Furthermore, in the case of conditionals where the branches join at some later point, the denotational semantics according to \cite{Zundorf2002} does not distinguish between the branches regarding rule applications.
This distinction, however, is essential to precisely define the concept of \emph{variable bindings}, i.e., matches of one story node that can be extended by subsequent story nodes. 
This, again, represents an operational semantic viewpoint and can be suitably handled by using \emph{scopes} as introduced in Section~\ref{sec:semantics}. 
For example, one might choose to allow bound variables within a branch but not after the branches join, regardless of if the conditional node was successful or not. 
This distinction and corresponding degrees of freedom in fixing such detailed design choices is not possible in the denotational semantics. 

The aforementioned constructs (sequential composition, conditionals, head-controlled loops) are all part of our basic SDM language and are described in the following Section~\ref{sec:semantics}, where we also recall the denotational semantics of each construct and discuss relevant differences in more detail.

%% file: po.tikz
\resizebox{2.2cm}{2cm}{
\begin{tikzpicture}[node distance=1.5cm, descr/.style={fill=white,inner sep=3pt}]
\node (L) {$ L$}; 
\node (K) [node distance=2cm,right of=L] {$ R$};
\node (ML)[below of=L]  {$G$}; 
\node (MK)[below of=K] {$H$};
\node (PO) at ($(L)!0.5!(MK)$) {$(PO)$};


\draw[->] (L) to node [descr,font=\footnotesize,swap]{$r$} (K);

\draw[->] (ML) to node [descr,font=\footnotesize,swap]{$r'$} (MK);

\draw[->] (L) to node [descr,font=\footnotesize]{$m$} (ML);
\draw[->] (K) to node [descr,font=\footnotesize,swap]{$m'$} (MK); 

%
%

\end{tikzpicture}
}

%% file: semantics.tex
\section{SDM Syntax and Semantics}
\label{sec:semantics}

In this section, we present two typed graph grammars, one for the syntax and one for the semantics of basic SDM. 
We specify the syntax graph grammar in a compact form (Sec.~\ref{sec:syntax}), as the primary focus of our paper is on the semantics. 
The syntax grammar generates syntactically valid control flows but does not specify the mapping of story patterns to story nodes. 
For the story patterns, we assume that they are properly typed over the input type graph and that all bindings are marked such that each bound variable has a previous occurrence in the story diagram where it was matched.

In Sections~\ref{sec:sdmmodel}--\ref{sec:ifelse}, we define the semantics of the basic SDM constructs on the basis of a type graph (Sec.~\ref{sec:sdmmodel}) also via standard GraTra rules.
We characterize our approach as a \emph{step semantics} in order to emphasize its focus on operational steps of the execution and to distinguish it from a fully-fledged SOS (where other concepts such as observable actions, labelling and equivalence mostly play an important role as well). 
It is important to define\begin{inparaenum}[(i)]
	\item how a \emph{state} of the system (here, a state of the story diagram execution) is characterized and
    \item which operations are explicitly included in the semantics, i.e., what is the notion of an observable \emph{step} in the semantics as expressed by the semantic rules.
\end{inparaenum} 
These notions are also covered in Section~\ref{sec:sdmmodel}.
Afterwards, we present the semantic rules by incrementally adding the following constructs to the already specified language fragment: initialization (Section~\ref{sec:init}), pattern matching (Section~\ref{sec:pm}), sequential chains of control flow nodes (Section~\ref{sec:seq}), as well as conditional control flow nodes and head-controlled loops (Section~\ref{sec:ifelse}).

\input{syntax}

\vspace{-0.4cm}
\subsection{Concept of the Semantics}\label{sec:sdmmodel}

\begin{figure}[!htbp]
	\centering
	\includegraphics[width=0.7\textwidth]{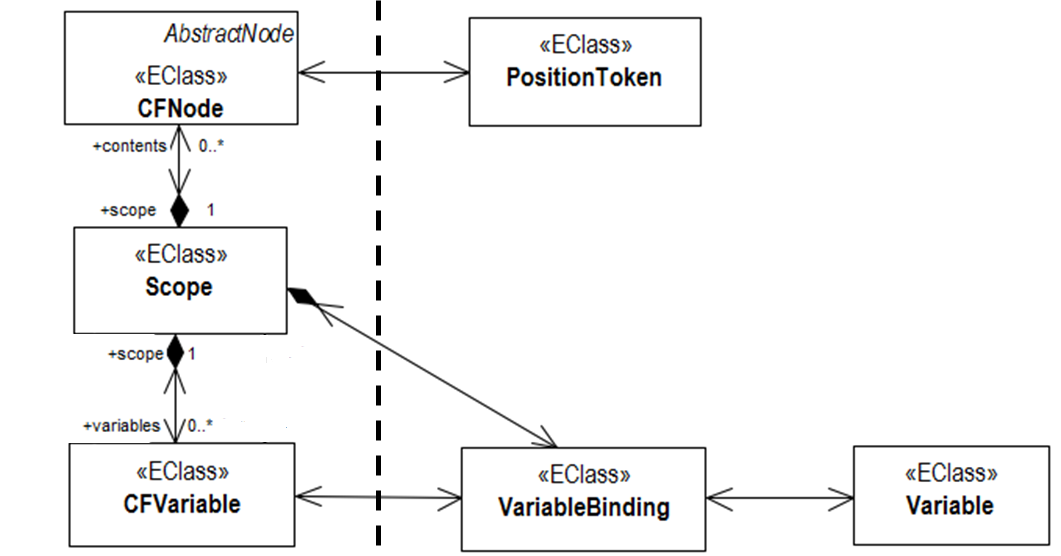}
	\caption{Type graph for SDM states}
	\label{fig:metamodel}
\end{figure}

Figure~\ref{fig:metamodel} depicts the type graph for story diagrams that we use to specify our semantics. 
The central elements of a story diagram are story nodes, modelled by the type \texttt{CFNode}. 
The class \texttt{CFNode} is also responsible for connecting the syntactic (Figure~\ref{fig:cfsyntax}) and the semantic type graphs (being a concept in both). 
\texttt{CFNode}s might be ordered into linked sequences (defining their execution order) using the next reference of the \texttt{AbstractNode} type (Figure~\ref{fig:cfsyntax}). 
\texttt{CFNode}s are contained in \texttt{Scope}s, as expressed by the containment arrow (a bidirectional arrow with a diamond on the container side). 
Basically, scopes represent the different blocks of the control flow and are ordered hierarchically. 
For instance, the scopes of conditional branches are sub-scopes of the \emph{parent scope}, where the branches originate from. 
There is always exactly one \emph{root scope} to which the first story node belongs.

A \texttt{Scope} contains control flow variables, which represent model elements. 
The correlation of these three elements (left of the dashed line) can be summarized as: the \texttt{CFVariable}s in a given \texttt{Scope} are those variables which are valid in the \texttt{CFNode}s of the same scope.

On the right-hand side of the dashed line, we can see the dynamic constructs used by the semantic rules. 
In each story diagram being executed, there is a single \texttt{PositionToken} that is attached to the current \texttt{CFNode} to be executed.
\texttt{VariableBinding}s represent the local variables of a scope, whose names are bound to an object in the input graph. 
This attachment is represented by a \texttt{VariableBinding} and a \texttt{Variable}, the latter representing our interface to the nodes in the input graph, which are not explicitly visible from the semantics. 
Thus, this triple (together with a \texttt{CFVariable}) captures a mapping $loc$ of local variables to input graph nodes $loc: CFVariable \rightarrow Variable$.

A graph typed according to Figure~\ref{fig:metamodel}, together with a control flow graph and story patterns constitute a valid \emph{state} of a story diagram execution. 
Such a valid state, together with the actual input graph (which might have already undergone some manipulations), constitutes a configuration of the story diagram execution. 
To avoid confusion, we stick to this terminology and refer to the semantic model of the story diagram execution as \emph{state}, to the input graph under manipulation further as model or input model, and to the pair of state and model as \emph{configuration}.
A \emph{step} of the semantics is a transition from one state to another which constitutes: \begin{inparaenum}[(i)]
	\item modifying the input model according to the pattern in the actual control flow node, 
	\item identifying the next control flow node and shifting the token there or terminating and
    \item optionally modifying other state elements such as bindings.
\end{inparaenum}

How such a step is actually specified is the main concern of the following sections. 
It is important to note that there is no one-to-one correspondence between GraTra rules and semantic steps. 
In our semantics, a semantic step consists of the application of multiple GraTra rules. 
All possible applications of these rules in a state constitute together one semantic step leading to the next state. 
As the possible applications are independent of each, their joint application and the next state are deterministic.

\textbf{Technical remarks.} The two type graphs presented (syntax and semantics) can be considered as a single one joined by the type \texttt{CFNode}.
We thus use types from both for specifying the rules. 
We assume that the scopes and control flow variables of a story diagram have been analysed and created before executing the semantics (hence scopes can be used as existing elements in the rules). 
This analysis and creation of scopes is possible on the control flow graph and the patterns in the control flow nodes without additional information; the corresponding algorithm to accomplish this is, however, out of scope.

\subsection{Initialization}\label{sec:init}

In this section, we define \emph{initialization} which, given a control flow graph with story patterns, creates the initial state for the semantics. 
In an initial state, the position token is set to the first control flow node after the start node and default bindings are added to the root scope. 
The resulting state, together with an unmodified input graph considered as user input, constitute the initial configuration for the execution.
We also interpret the necessary preparations of creating the fixed \texttt{this} variable as part of the initialization.

\begin{figure}[!htbp]
	\includegraphics[width=\textwidth]{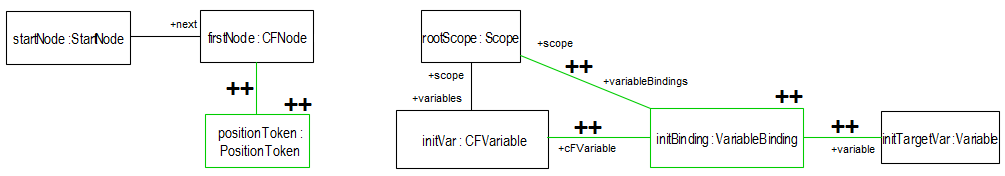}
	\caption{Initialization}
	\label{fig:initsemantics}
\end{figure}

\smallskip
\noindent\textbf{Semantics.} Figure~\ref{fig:initsemantics} depicts the initialization semantics consisting of two rules. 
In the left rule (applied exactly once), the position token is created and set on the first control flow node (the one directly connected to the start node in the control flow graph). 
In the right rule, for the initial variables in the root scope (in basic SDM, only for \texttt{this}), the bindings are created and also attached to the root scope. Note that initialization does not have an analogue in the higher-level denotational approach~\cite{Zundorf2002} as there is no need to handle these operational details.

\subsection{Pattern Matching}\label{sec:pm}

In this section, we consider a single control flow node with a story pattern. 
When the execution of the story diagram proceeds to a story node, its story pattern is executed.

\noindent\textbf{Semantics.} The semantics of pattern matching requires special handling and follows the same principles for our semantics and the denotational approach~\cite{Zundorf2002}, as both rely on GraTra rule applications. 
Although our step semantics focuses on the control flow, the actual model transformation is still performed by the story patterns in the control flow nodes. 
As soon the position token is set to point to a new control flow node, the matching of the pattern contained in the control flow node is performed. 
There are two cases possible: a match can be found and model objects become bound via the variable names in the pattern, or no match can be found. 
In the former case, new variable bindings are created (unbound black and green elements) and some are deleted (red elements); in the latter case, a failure is reported. 

We abstract away the actual matching and rule application process (which is performed according to the SPO approach), as from a semantic point of view, we are only interested in the result, i.e., which values (model objects) are attached to control flow variables. 
Moreover, we do not explicitly include the modifications which the input model undergoes in case of a successful match; we consider this step as a part of the pattern matching process and we assume that we get information on which variables are newly bound and which have to be deleted.
%
To capture the effects of pattern matching in our semantics, we introduce a new type \texttt{Pattern\-Invocation}. 
It is contained by the control flow node and represents the result of the pattern matching process regarding variables, i.e., which unbound variables become bound and which bindings get deleted. 

\begin{figure}[!htbp]
	\centering
	\includegraphics[width=\textwidth]{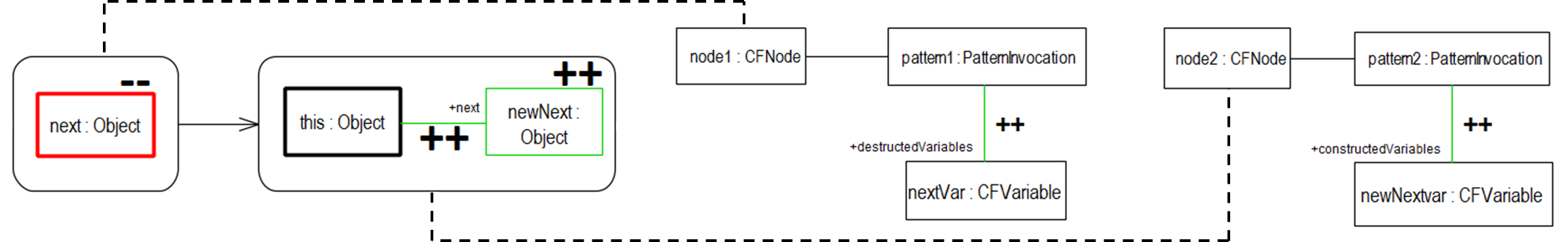}
	\caption{Result of pattern executions}
	\label{fig:pmsemexample}
\end{figure}

\noindent\textbf{Example.} Figure~\ref{fig:pmsemexample} depicts two subsequent control flow nodes from Figure~\ref{fig:example_dno3} on the left hand side, and a corresponding representation of how the results of the rule applications appear in the semantic model. 
The effect of the deletion in the first control flow node is shown in the middle, where the corresponding variable is added to the pattern invocation \texttt{pattern1} with a \emph{destructedVariables} reference. Creating a new object in the second control flow node is shown on the right, where the corresponding variable is added to \texttt{pattern2} with a \emph{constructedVariables} reference. Note that deletion and creation might also be combined in a pattern and there might be multiple variables to delete/create.

\subsection{Sequential Chain of Story Nodes}\label{sec:seq}

In this section, we consider simple, sequential chains of control flow nodes, which are executed in a strict order and where it is not allowed to create loops in the chain.
The chain might be arbitrarily long, but it is strictly sequential, i.e., it contains no loops. 
In a complete, valid story diagram where no other construct is allowed (i.e., a specification in the SDM sub-language that consists only of sequential chains), the chain starts with a start node and ends with a stop node. 
From a semantic point of view, we regard the stop node as a control flow node having no pattern and no outgoing arrow. 
When the position token arrives at such a node, no rule fires any more and the execution terminates.
%

\smallskip
\noindent\textbf{Semantics.} 
Denotationally~\cite{Zundorf2002}, the semantics of two subsequent control flow nodes $n_1$ and $n_2$ with rules $r_1$ and $r_2$, respectively, is given in terms of input and output graphs as $Sem(n_1,n_2) = \{(G,H) | \exists G': G \overset{r_1}{\Longrightarrow} G' \overset{r_2}{\Longrightarrow} H \}$ (longer chains are defined by induction). The most important difference compared to a more detailed step semantics is that bindings are not handled explicitly. 
The semantic definition of sequential composition only requires that there is \emph{an} intermediate graph between the two applications, but does not assign the resulting pairs to their respective matches. As mentioned before, this might be an appropriate level of abstraction for testing and verification tasks but might be too high-level for tool developers.

\begin{figure}[!hbtp]
\centering
	\includegraphics[width=\textwidth]{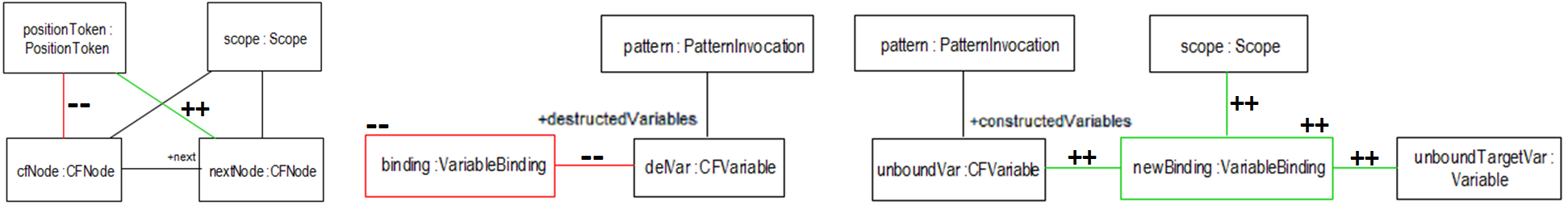}
	\caption{\emph{seqSuccess}: token shift and binding updates after successful pattern matching}
	\label{fig:seqsemantics}
\end{figure}

Regarding our step semantics, subsequent control flow nodes in a sequential chain all belong to the same scope. 
A step in our semantics corresponds to a shift of the position token from the current control flow node to the next one if pattern matching was successful. 
We also have to handle the case where no match for the pattern can be found. 
In basic SDM, this case is handled by terminating with an error. 
We model this error by detaching the position token from the control flow to clearly distinguish between normal and erroneous states, as otherwise, the position token is always attached to a control flow node. 
We do not show graphically the fairly trivial rule of deleting the edge between the position token and the actual control flow node. 
The step for handling a match is depicted in Figure~\ref{fig:seqsemantics}. 
Note that the binding update rules (middle and right ones) are executed for every relevant variable correspondingly marked in the pattern invocation step.

\noindent\textbf{Example.} Considering the two subsequent control flow nodes on the left of Figure~\ref{fig:pmsemexample}, after the first node has been executed, the position token is passed to the second one and the binding of nextVar gets deleted, corresponding to the middle rule in Figure~\ref{fig:seqsemantics}.

\subsection{Conditional Control Flow Nodes and Head-Controlled Loops}\label{sec:ifelse}

In this section, we extend our language by a conditional construct. 
\emph{Conditional control flow nodes} have exactly two outgoing arrows leading to other control flow nodes, one labelled with \emph{Success}, and the other with \emph{Failure}. 
In the case of a successful matching process, the \emph{Success} edge will be followed, otherwise the \emph{Failure} edge is taken. 
It is important to note that the notion of conditionals in SDM is not completely analogous to the traditional conditional statements of imperative programming: choosing the next block to be executed does not depend on evaluating a boolean expression over already bound variables, but depends rather on the result of a pattern matching process, during which new variables might become bound and existing bindings deleted.

A special form of a conditional construct is when one of the branches returns to the conditional node, resulting in a head-controlled loop, also simply known as a \emph{while} loop. After the recurring branch has been executed, the control is passed to the conditional node again, taking the possibly changed model as input; the conditional is then re-evaluated independently of the previous iteration, just as in a traditional imperative language.
%
Another special case that must be explicitly handled is when  two branches \emph{join} again at a control flow node.
We shall discuss different strategies of handling variable bindings.


\smallskip
\noindent\textbf{Semantics.} In the denotational approach~\cite{Zundorf2002}, the semantics of a conditional $\textit{if } n_1 \textit{ then } n_2 \textit{ else } n_3$ (where $n_1, n_2, n_3$ are control flow nodes with rules $r_1, r_2, r_3$, respectively) is: $Sem_{if}(n_1,n_2,n_3) = Sem(n_1,n_2)$ if $\exists\, G': G \overset{r_1}{\Longrightarrow} G'$ with input $G$, and $Sem(n_1,n_3)$ otherwise.
Note that it is assumed in this semantics that SDM execution always continues even after a failing match. 
This does not allow for the distinction that in the case of sequential nodes, a failing match should result in abrupt termination while in the case of a conditional, execution  continues along the \emph{Failure} branch. 
Our approach handles both cases explicitly.

Denotationally, the semantics of a \emph{while} loop: $\textit{while } n_1 \textit{ do } n_2$, with control flow nodes $n_1, n_2$, rules $r_1,r_2$, and input graph $G$ is given as $Sem_{while}(n_1,n_2) = Sem(n_1,n_2,\textit{while } n_1 \textit{ do } n_2)$ if $\exists G': G \overset{r_1}{\Longrightarrow} G'$ and $Sem(n_1)$ otherwise. Note that this specification only handles looping along success, whereas our step-based approach can also handle a ``negative loop'' where the recurring branch goes along failure.

Regarding our step semantics, in order to define the relations between the conditional node and subsequent nodes, we make use of an additional reference type between two control flow nodes: \emph{success} and \emph{failure} (as shown in Figure~\ref{fig:cfsyntax}).
A more significant difference is that in the case of conditionals, we now utilize different scopes. 
Both branches of a conditional have their own separate scopes, whose \emph{parent scope} is the scope of the conditional node.
The branch scopes get dynamically created as execution proceeds to them for the purpose of handling \emph{while} loops. 
Deciding which bindings they should inherit and pass on to their parent scope is an interesting design decision, especially if the branches join at some later point; we elaborate on this subject to demonstrate the capabilities of our approach as well as to emphasize possible alternative semantics and ambiguities.

As soon as the execution of a branch of a conditional is completed, i.e., execution of a node that is reachable from both branches of the conditional commences, we revert to the parent scope (the scope of the conditional node). 
In the conditional node itself or somewhere in one of the branches, we might, however, have updated some bindings. 
The case of variables that are newly bound (or created) \emph{in only one branch} is clear: such bindings must lose their validity when the branch scope is exited (as is the case in standard imperative languages), and are not passed on to the parent scope.
For variables that are newly bound or created \emph{in both branches}, one could take different approaches for cases where this can be guaranteed statically.
We suggest a conservative approach where such bindings also lose their validity, i.e., identical variable names do not necessarily represent the same variable.
An optimistic approach, however, where such variables do become available after the branches are merged also makes sense, implying that the conditional represents different ways to bind essentially the same variable. 

The case of deletion is equally interesting: If a variable is already bound before the conditional node, it would also be available after the branches have joined as these parts of the story diagram belong to the same scope. 
If this variable is, however, deleted \emph{in any} of the branches, this could result in invalid bindings in the parent scope.
We suggest again a conservative approach, where such bindings are removed from the parent scope, as opposed to an optimistic approach, where such bindings are retained, possibly leading to failures at runtime.

\begin{figure}[!htbp]
	\includegraphics[width=\textwidth]{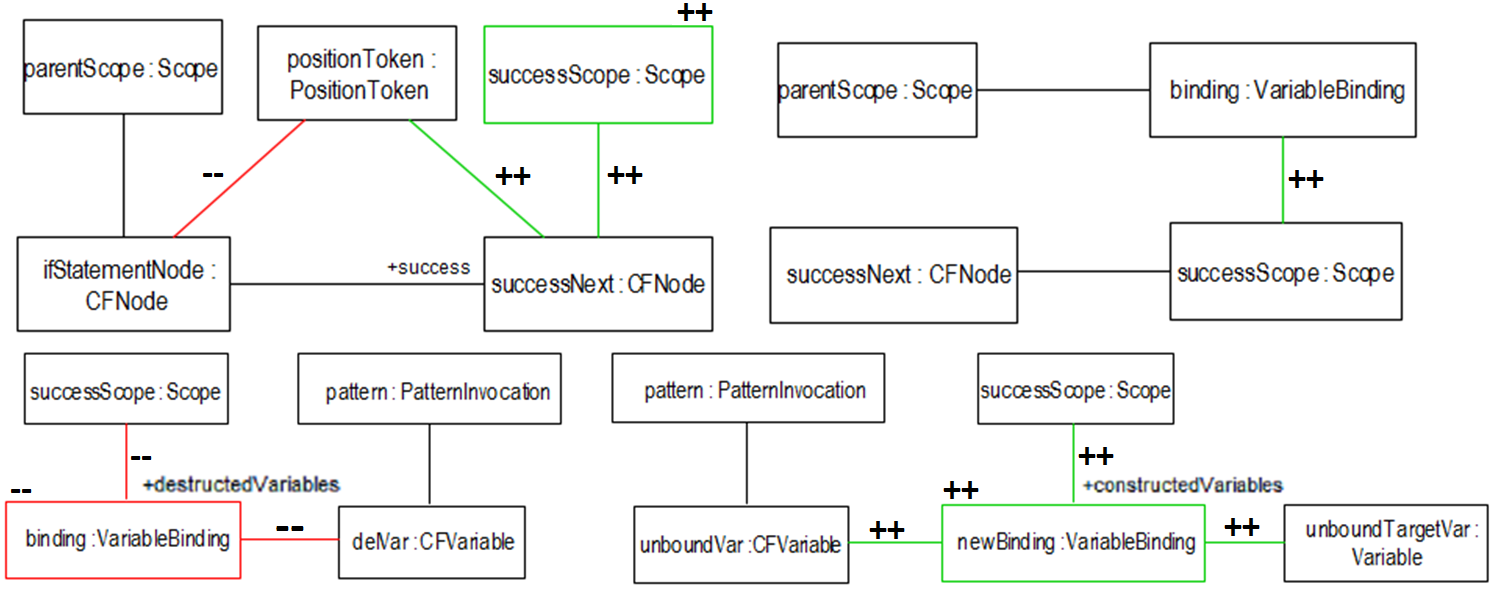}
	\caption{Handling a match of the conditional node}
	\label{fig:ifelsesemantics}
\end{figure}

Figure~\ref{fig:ifelsesemantics} depicts the rules for handling a conditional with a successful match (i.e., the \emph{Success} branch). 
Note that all the objects, particularly the scopes, are uniquely determined as they are attached to their respective control flow nodes whose connection is unique through the \emph{success} reference. 
The upper left rule creates the scope for the success branch and passes the token to its first node. 
In the other rules, \texttt{successScope} always denotes this scope just created. 
The upper right rule copies the bindings of the parent scope to the fresh sub-scope. 
The binding updates of the conditional node (rules in the bottom row) are then performed in the sub-scope \texttt{successScope}.
Note that the upper rules can be used for handling an unsuccessful matching in a conditional if we substitute failure for success in each relevant element (all bindings are inherited from the parent scope, token is shifted, but no bindings are updated). 
Creating a fresh scope for conditional branches upon entering them is crucial for handling \emph{while} loops, where one of the branches returns to the conditional node.
In this case, the pattern in the conditional node gets matched again and we proceed accordingly. 
Creating a fresh scope each time prevents bindings in branch scopes from appearing in subsequent iterations.

\vspace{-0.3cm}
\vspace{-0.3cm}

The ``conservative'' rule for handling deletion inside a conditional branch is depicted for the success branch in the figure to the right below (analogously for failure). 
It specifies that all variables bound in the parent scope, for which there is no binding in the success scope (NACs are represented by crossed-out edges), must be invalidated in the parent scope, making these bindings unavailable after the join.

\begin{wrapfigure}{hL}{0.45\textwidth}
	\vspace{-0.55cm}
	\includegraphics[width=0.44\textwidth]{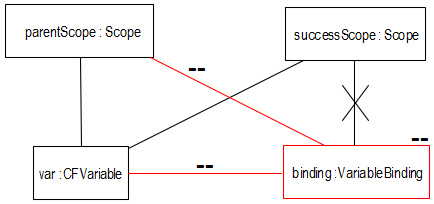}
	\label{fig:conddel}
\end{wrapfigure}
Specifying an ``optimistic'' rule to make all elements matched in \emph{both} branches available as bound in the parent scope after the join is fairly straightforward. 
Although we do not specify this rule and other possible variants in the present paper, the aforementioned considerations demonstrate the advantages of an explicit step-based semantic approach in clarifying such details.
\clearpage

\noindent\textbf{Example.} The bottom row in Figure~\ref{fig:example_dno3} represents a conditional where the branches join. If the match on the left is successful, a binding for \texttt{next} is created which is used in the \emph{Success} branch to delete this very object. If there is no match for \texttt{next}, we directly move on to the join node on the right. 
According to our semantics, this node belongs again to the same scope as the conditional, which means that there is definitely no valid binding for \texttt{next}.

%% file: syntax.tex
\subsection{Control Flow Syntax}\label{sec:syntax}

In this section, we define a graph grammar $GG_{Syntax} = (G_s,\mathsf{R}_{Syntax})$, which generates the set of all valid control flow graphs of basic SDM. 
According to this specification technique, a control flow is considered as a graph (also called \emph{control flow graph}) that represents the imperative control structure of a story diagram (without considering story patterns). 
We define $GG_{Syntax}$ by specifying a \emph{start graph} $G_s$ (considered as the minimal valid control flow) and a set $\mathsf{R}_{Syntax}$ of GraTra rules which specify the possible expansions of the start graph resulting in valid control flow graphs. 
To guarantee the validity of the resulting control flow graphs, we specify the rules in a way that every possible rule application to a valid control flow graph yields also a valid one.

\begin{figure}[!htbp]
	\centering
	\includegraphics[width=0.56\textwidth]{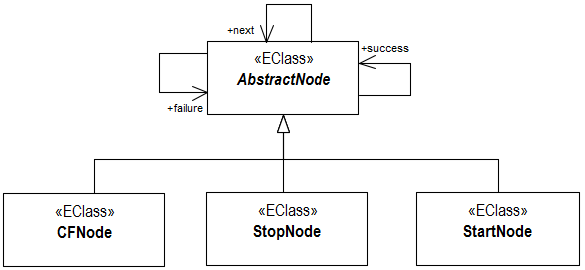}
	\caption{Type graph $TG_{Syntax}$ of the control flow syntax graph grammar $GG_{Syntax}$}
	\label{fig:cfsyntax}
\end{figure}

Both the start graph and the GraTra rules are typed over the type graph $TG_{Syntax}$, depicted in Figure~\ref{fig:cfsyntax}. 
The types \texttt{CFNode}, \texttt{StopNode} and \texttt{StartNode} have a common parent type, \texttt{AbstractNode}, from which they inherit all the edge types. 
These edge types are: \emph{next} for sequences of control flow nodes, as well as \emph{success} and \emph{failure} for the two branches of conditional nodes. 
All the edge types are represented as self-edges of the abstract type \texttt{AbstractNode} as such edges in a control flow run from one control flow node to another.
Introducing this abstract node enables a simplified, compact syntax grammar.

\begin{figure}[!htbp]
	\begin{minipage}{0.6\textwidth}
		\centering
		\includegraphics[width=\textwidth]{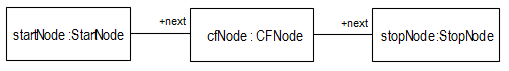}
		\captionof{figure}{Start graph $G_s$ of $GG_{Syntax}$}
		\label{fig:startgraph}
	\end{minipage}
	\hspace{0.02\textwidth}
	\begin{minipage}{0.38\textwidth}
		\centering
		\includegraphics[width=\textwidth]{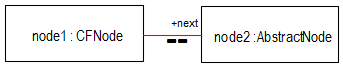}
		\captionof{figure}{The left-hand side $L$ shared by all rules in Table~\ref{table:syntax}}
		\label{fig:syntaxLHS}
	\end{minipage}
\end{figure}

Figure~\ref{fig:startgraph} depicts the start graph $G_s$ which is, consequently, the minimal valid control flow graph. 
$G_s$ consists of a sequence of a start node, a story node, and a stop node. 
Table~\ref{table:syntax} shows the rules in $\mathsf{R}_{Syntax}$ for the success case. 
For all rows apart from the first and the second, there are additional, completely symmetric rules for each one depicted, formed by substituting failure for success and vice versa, as success and failure are dual concepts. 
We also distinguish between different branch configurations as seen, e.g., in the third and fourth rows. 
We merge these two possibilities into a single row in the last two cases due to space restrictions. 
The rules are defined according to the standard GraTra notions of left-hand side ($L$) and right-hand side ($R$) graphs as introduced in Section~\ref{sec:gratrabasics} -- nevertheless, as each rule shares a single common $L$ graph, we do not include it for each rule but only specify it once in Figure~\ref{fig:syntaxLHS}. 
For the sake of clarity, we additionally use the SDM visualization of deletion and creation and mark the elements to be deleted in $L$ with red (resp. \texttt{-{}-}) and the elements to be created in $R$ with green (resp. \texttt{++}).
The rules depicted in Table~\ref{table:syntax} only give rise to well-formed, i.e., valid control flow graphs. We start from a minimal, but complete control flow graph $G_s$. 
In each step, we either insert a simple sequential story node or introduce a conditional, possibly with a loop. Conditionals without loops are handled in rows 2-4 and \emph{while} loops in rows 5-7. Considering also the symmetric rules not depicted here, $\mathsf{R}_{Syntax}$ consists of 16 rules: 1 for sequential composition, 1 for joining conditional branches, 4 for non-joining conditional branches and 10 for \emph{while} loops.

\vspace{-0.4cm}

\begin{table}
	
\begin{center}
	
\begin{tabular}{| b{2.8cm} | c|}
	
	\hline
	Description & Right-hand side of syntactic rule (common left-hand side in Figure~\ref{fig:syntaxLHS}) \\
	
	\hline
	Inserting a node sequentially & \includegraphics[trim=0 0 0 -5, width=0.5\textwidth]{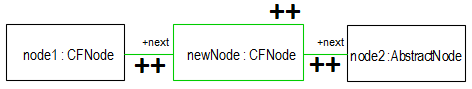} \\
	
	\hline
	Conditional with joining branches & \includegraphics[trim=0 0 0 -5,width=0.5\textwidth]{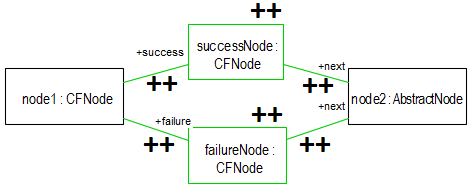} \\
	
	\hline
	Conditional with opening up a \emph{Failure} branch containing only a stop node & \includegraphics[trim=0 0 0 -5,width=0.5\textwidth]{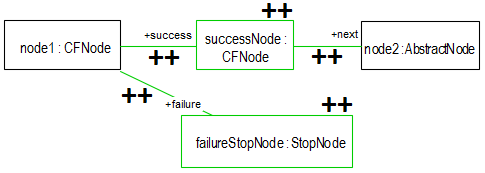} \\
	
	\hline
	Conditional with opening up a \emph{Failure} branch containing a control flow node and a stop node & \includegraphics[trim=0 0 0 -5,width=0.5\textwidth]{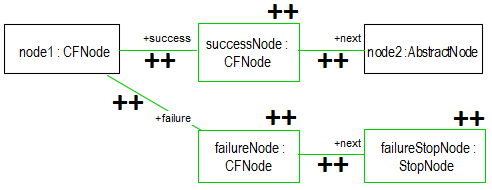} \\
	
%

	\hline
	Direct loop along \emph{Success} & 
	\includegraphics[trim=0 0 0 -5,width=0.5\textwidth]{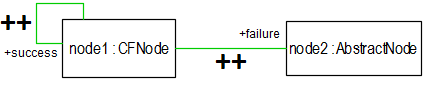} \\
	
	\hline
	Loop consisting of two nodes along \emph{Success} (both \emph{Failure} variants)  & 
	\includegraphics[trim=0 0 0 -5,width=0.8\textwidth]{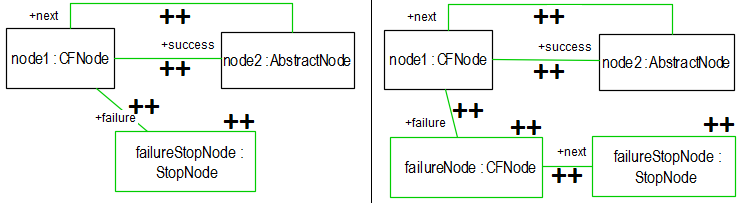}\\
	
%
	\hline
	Loop consisting of three nodes along \emph{Success} (both \emph{Failure} variants) & 
	\includegraphics[trim=0 0 0 -5,width=0.8\textwidth]{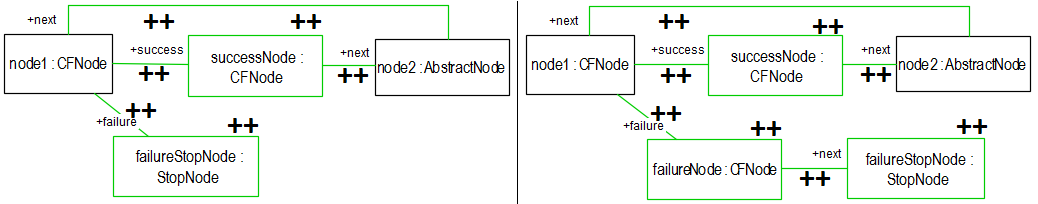}  \\
	
%
	\hline
	
\end{tabular}
\end{center}
\caption{Graph grammar for control flow syntax}
\label{table:syntax}
\end{table}

%% file: conclusion.tex
\section{Conclusion and Future Work}
\label{sec:conclusion}

In this paper, we have presented a novel specification approach for a step semantics of SDM, where the steps correspond to GraTra rule applications in a provided graph grammar. 
Although we only examined basic SDM constructs, we have demonstrated the potential of our specification technique by presenting an alternative to a state-of-the-art semantics and by discussing possible design decisions (conservative vs. optimistic) concerning allowed variable bindings before and after conditionals (and while loops).

As also illustrated by our work, there is no definitive approach for defining language semantics and a detailed, step-based semantics can complement a high-level, denotational one. The domain of the latter does not include anything more than graphs resulting from a chain of rule applications, which makes it suitable to analyse standard GraTra notions such as confluence in a programmed scenario. 
However, our complementary approach is tailored to guide and facilitate providing programmed GraTra tooling in practice, where details hidden in the denotational semantics often play a crucial role. 
The price we pay is a substantially more elaborate machinery, dealing with finer details of SDM execution.

Nonetheless, our initial results presented in this paper have proven that the size of the semantic specification remains manageable even in the case of conditionals and loops; compound constructs can be specified by reusing parts of simpler steps. 
We do not expect the complexity of the semantic steps to explode when considering further constructs.
We have experimentally implemented the semantic rules from this paper in the graph transformation tool eMoflon, realizing the rules on top of the actual SDM meta-model of eMoflon.
This is promising regarding the practical applicability of our approach, e.g., for static analysis or refactoring. 

The most obvious and immediate item of future work is to expand the semantics to cover further SDM language constructs such as SDM method calls and \emph{for-each} style loops.
These extensions can be based on the semantics presented in this paper. 
Another promising field of future investigation is extending SDM with a notion of multi-threading, where the single threads are SDM method executions. 
As a relevant application area, research on graph-based networking could profit from such an extension in various ways.

From a theoretical point of view, we plan to investigate the possibilities which arise from using a semantic style that is already close to well-known \emph{operational} semantics. 
In particular, we plan to elaborate on a \emph{trace} notion for SDM based on the defined steps and transition labelling based on GraTra rule applications. 
These concepts would enable us to leverage advanced techniques such as defining a suitable equivalence concept for SDM.
